\documentstyle[12pt]{article}
\begin{document}
\setlength{\topmargin}{-4em}
\addtolength{\textheight}{4cm}
\title{Spontaneous Symmetry Breaking Mechanism in Light-Front Quantized Field
Theory- (Discretized Formulation)\thanks{\it AIP Conference
 Proceedings 272, ed. J. Sanford, XXVI
Int. Conf. on High Energy Physics, Dallas, Texas, August 1992, pg. 2125 . }}
\author{\em Prem P. Srivastava\thanks{Permanent address: Centro Brasileiro de
Pesquisas F\'{\i}sicas, Rio de Janeiro, RJ, Brasil}\\
\cr
Department of Physics\\
The Ohio State University, \\
Columbus, OH, U.S.A.}
\date{ }
\maketitle

\begin{abstract}
The scalar field is quantized in the discretized light-front framework
following the {\em standard} Dirac procedure and its infinite volume limit
taken.
The background field and the nonzero mode variables do not commute for finite
volume; they do so only in the continuum limit. A {\em non-local constraint}
in the theory relating the two is shown to follow and we must deal with it
along with the Hamiltonian. At the tree level the constraint leads to a
description of the spontaneous symmetry breaking. The elimination of the
constraint would lead to a highly involved light-front. Hamiltonian in contrast
to the one found when we ignore altogether the background field. The
renormalized constraint equation would also account for the instability of the
symmetric phase for large enough couping constant.
\end{abstract}

\vspace{1cm}
\noindent
SLAC (HEP-) Database: PPF-9222, April 1992.
\newpage
\section*{ }

{\bf 1.} The possibility of building a Hamiltonian formulation of relativistic
dynamics on light-front surface, $\tau =(t+z)=const.$, was pointed out by Dirac
\cite{Dirac} and rediscovered by Weinberg \cite{Wein} in the context of
old-fashioned perturbation theory in the infinite-momentum frame. Since the
longitudinal momentum $k^+$ (in the massive case) is necessarily positive and
conserved, the vacuum structure looks simpler. The discretized light-cone
quantized (DLCQ) field theory \cite{Pauli} in the context of perturbation
theory does show simplifications and one hopes that the non-perturbative
calculations may also be manageable through the numerical computation. The
recent developments in the studies on Light-front Tamm-Dancoff Field Theory
\cite{Wil} to study non-perturbative effects (e.g., the relativistic bound
states of light fermions) and the begining of a systematic study of
perturbative renormalization theory \cite{Mus} have the same motivation. The
light-front approach may as described in ref. \cite{Wil} throw some light on
the relationship between the constituent quark picture and QCD with the dense
sea of quarks and gluons in its vacuum.

The description of non-perturbative vacuum structure, for example, in the
presence of a spontaneuos symmetry breaking scalar potential, Higgs mechanism,
the fermionic condensates, and other related problems have, however, remained
in the light-front framework without a clear understanding even at the tree
level (see reviews in \cite{Pauli,Wil}) in contrast to the description
available
in the case of equal-time quantization. We consider here, for concreteness and
since it reveals most of the essential points, the light-front quantization of
the massive scalar field in $1+1$ dimensions in detail. We will work for
convenience in the discretized formulation and recover latter the physically
significant predictions by taking the infinite volume limit to remove the
spurious contributions introduced, for example, through the (zero modes) of
finite volume delta and epsilon functions. Since the Lagrangian of any theory
written in terms of light-front coordinates is necessarily singular (or
degenerate), some new features emerge compared to the case of the equal-time
quantization. For example, the background field variable $\omega$ and the field
variable $\varphi$ describing the fluctuations above the background may not be
quantized independently and they satisfy a {\em nonlinear constraint} equation.
Moreover their commutator is vanishing only in the continuum limit; in the
discretized formulation with finite volume it does not. These features have
been overlooked in DLCQ where the background field is ignored altogether and we
work with a very simple Hamiltonian, simple commutation relations among the
creation and destruction operators, and a trivial vacuum. This is {\em not} the
case if non-perturbative $(\omega \neq 0)$ effects are taken into account. We
show below that in the light-front quantized theory we must now deal not only
with the Hamiltonian operator but also with a non-linear operator constraint.
It is not practical to solve the latter since it will lead to a very involved
Hamiltonian unlike the simple one usually adopted in the literature
\cite{Pauli} ignoring the background field. It follows from the discussion
below that the constraints are very useful ingredients and evidently may not be
ignored in the light-front quantized field theory when considering
non-perturbative effects.  In the case under discussion it allows us to obtain
the usual criterian for the spontaneous symmetry breaking at the tree level;
there is no physical argument to minimize the light-front energy in contrast to
the equal-time case. The renormalization may also be handle and the
renormalized constraint equation may be used to obtain the quantum corrections
in the value of the background field. It is worth remarking that this may be
achieved in a straightforward fashion only in the continuum limit where
$\omega$ and $\varphi$ now commute. We comment also on the description of the
well known phase transition for large coupling constants \cite{Chang} and high
order quantum correction to the well known ``duality'' \cite{Col} relation.

The Hamiltonian formulation for singular Lagrangians may be obtained using the
Dirac method \cite{Di} which also allows us to find the constraints and the
modified (Poisson) brackets necessary to take care of them among the surviving
dynamical variables. Some constraints may of course be read off even at the
Lagrangian level, say, by integrating the equations of motion but to quantize
the theory we need to build a canonical framework. In the present context the
Dirac method was attempted in the refs. \cite{Mask,Th} in incomplete fashion
with inconclusive results. The {\em standard} Dirac method requires that {\em
all} the constraints (except the gauge-fixing ones) are derived from within the
given Lagrangian. In ref. \cite{Mask} the constraint $p\approx 0$ discussed
below was missed while in ref. \cite{Th} it was proposed to modify the
procedure itself and constrainsts added to the theory from outside. The authors
also do not take the continuum limit properly, suggest modifications to the
well known light-front commutators, do not consider the full implications of
the constraint arising in the original Dirac procedure. There is also some
misunderstanding about the term ``zero mode''. We have as the candidates the
background field variable $\omega$ (e.g., the vacuum expectation value of the
scalar field $\phi$) and the variable $\stackrel{\sim}{\varphi}(0)$ -- the
Fourier coefficient (or transform) for $k^+=0$ of the field $\varphi$. The
kinematical constraint in the light-front frame work for massive theories
requires that we must set $\stackrel{\sim}{\varphi}(0)$ to be vanishing since,
it corresponds to a {\em space-like} momentum vector $k^{\mu}$ in contrast to
the nonzero (longitudinal) modes which correspond to the {\em time-like}
vectors. This conclusion is also supported from axiomatic field theory
considerations \cite{S}. The ref. \cite{Mask} does not even mention the
background field $\omega$ which in the well known equal-time case characterizes
the non-perturbative vacua when dealing with the spontaneously bronken symmetry
potential while $\stackrel{\sim}{\varphi}(0)$ is taken to be nonvanishing. Once
clarified on these points and all the constraints derived from inside the
theory, it is shown here that there is no need to modify the Dirac method
thereby
casting doubts on the validity of the Dirac method in general. It becomes clear
(Sec. 2) also that there is no problem as regards to taking the infinite volume
limit in straightforward way of the discretized formulation. In our context the
constraints were not noted in still earlier work \cite{J}, where only the
light-front commutation relations were derived using the Schwinger's
variational principle. We show here that they also emerge from the Dirac method
and are accompanied by the (Hamiltonian and) nonlinear constraints. Similar
constraints in the renormalized QCD in the light-front framework through a
perturbative expansion may give useful hints to find new counter terms.

\section*{ }
{\bf 2.} The light-front Lagrangian for the scalar field $\phi$ is
\begin{equation}
\int_{-\infty}^{\infty}dx [\dot{\phi}\phi'-V(\phi )],
\end{equation}
where $V(\phi )\geq 0$, for example, $V(\phi )=(\lambda /4)(\phi^2-m^2/\lambda
)^2$, the potential with the wrong sign for the mass term and $\lambda \geq
0,\;
m\neq 0$. Here an overdot and a prime indicate the partial derivations with
respect to the light-front coordinates $\tau \equiv x^+=(x^0+x^1)/\sqrt{2}$ and
$x\equiv x^-=(x^0-x^1)/\sqrt{2}$ respectively and $x^+=x_-,x_+=x^-$ while
$d^2x=d\tau dx$. The Euler equation of motion, $\dot{\phi}'=(-1/2)V'(\phi )$,
where a prime on $V$ indicates the variational derivative with respect to
$\phi$, shows that classical solutions, for instance, $\phi$=const., are
possible to obtain. We separate from the classical field $\phi$
the variable $\omega =\omega (\tau )$
(background field) corresponding to the operator which gives the vacuum
expectation value of the would be quantized field $\phi$. The {\em generalized}
function $\phi$, is then expressed as $\phi (x,\tau )=\omega (\tau )+\varphi
(x, \tau )$. Here $\varphi$ is an ordinary absolutely integrable function of
$x$ such that its Fourier transform $\stackrel{\sim}{\varphi}(k,\tau )$ (or
series) exist together with the inverse transform. Since under these
assumptions $\varphi \rightarrow 0$ for $|x|\rightarrow \infty$ the vacuum
expectation value of the quantized field $\varphi$ vanishes. The vacuum
expectation values of the momentum space operators $\stackrel{\sim}{\varphi}(k,
\tau )$ then also vanish for all $k\equiv k^+$. The quantized theory vacuum
will be defined such that the expectation value of these operators for $k$
different from zero is vanishing and we will assume the absence of the zero
mode in $\varphi$, e.g., $\stackrel{\sim}{\varphi}(0)=0$ so that the space
integral of the $\varphi$ vanishes. This follows from the kinematical
constraint in massive theory which allows only $k>0$ as pointed out in Sec. 1.
It is also clear that the discretized version may also be constructed and its
infinite volume limit taken without any problems. The final expressions in the
alternative treatments do coincide in the continuum limit if we adopt the
commonly used interpretation of the space integrals over $[-\infty ,\infty ]$
as integral over $[-L/2,L/2]$ when $L\rightarrow \infty$ (Cauchy principal
value) and for the delta function with vanishing argument, $2\pi \delta
(0)=\lim_{L\rightarrow \infty}\int_{-L/2}^{L/2}dx$, where $L$ is to be
identified with the finite size extension in the $x$ direction in the
discretized formulation.

The discretized formulation is obtained through the following Fourier series
expansion
\begin{equation}
\phi (\tau, x) = \frac{q_0}{\sqrt{L}} + \frac{1}{\sqrt{L}}\sum_n{}'\; q_n(\tau
)
e^{-ik_nx}\equiv \frac{q_0}{\sqrt{L}}+\varphi(\tau ,x)
\end{equation}
where periodic boundry conditions are assumed, $\Delta =(2\pi /L),k_n=n\Delta ,
\;\linebreak n=0,\; n=0,\pm 1, \pm 2,\cdots ,\; \sum{}'_n$ indicates the
summation excluding $n=0$. It is only {\em for convenience} that we have
introduce an explicit factor $1/\sqrt{L}$ in the first term of (2); its limit
in the continuum is simply $\omega (\tau )$. The discretized Lagrangian
obtained by integrating the Lagrangian density in (1) over the finite interval
$-L/2\leq x\leq L/2$ is given by
\begin{equation}
i\sum_n\; k_nq_{-n}\; \dot{q}_n \; -\; \int_{-L/2}^{L/2}dx\; V(\phi )
\end{equation}
The momenta conjugate to $q_n$ are $p_n=ik_nq_{-n}$ and the canonical
Hamiltonian is found
to be
\begin{equation}
H_c = \int_{-L/2}^{L/2}dx \; V(\phi )
\end{equation}
The primary constraints are thus $p_0\approx 0$ and $\Phi_n\equiv
p_n-ik_nq_{-n}\approx
0$ for $n\neq 0$. We postulate initially the standard Poisson brackets at equal
$\tau , viz, \{p_m,q_n\}=-\delta_{mn}$ and define the preliminary Hamiltonian
\begin{equation}
H' = H_c+\sum_n{}'\; u_n\Phi_n + u_0p_0
\end{equation}
On requiring the persistency in $\tau$ of these constraints we find the
following weak equality \cite{Col} relations
\begin{eqnarray}
\dot{p}_0 &=& \{p_0,H'\} = \{p_0,H_c\} = \frac{1}{\sqrt{L}}\int_{-L/2}^{L/2} dx
V'(\phi ) \equiv -\frac{1}{\sqrt{L}}\; \beta(\tau ) \approx 0\; , \\
\dot{\Phi}_n &=& \{\Phi_n,H'\} = -2i\sum_n{}'\; k_n\; u_{-n}\;
-\frac{1}{\sqrt{L}} \int_{-L/2}^{L/2}dx\; V'(\phi )e^{-ik_nx}\approx 0.
\end{eqnarray}
{}From (6) we obtain an interaction dependent secondary constraint $\beta
\approx
0$ while (7) is a consistency requirement for determining $u_n,\; n\neq 0$.
Next we extend the Hamiltonian to
\begin{equation}
H''=H'+\nu (\tau )\beta (\tau )\; ,
\end{equation}
and check again the persistency of all the constraints encountered above making
use of $H''$. We check that no more secondary constraints are generated if we
set
$\nu \approx 0$ and we are left only with consistency requirements for
determing the multipliers $u_n,\;u_0$. We easily verify that all the
constraints
$p_0\approx 0,\; \beta \approx 0$, and $\Phi_n\approx 0$ for $n\neq 0$ in our
system are second class \cite{Col}. They may be implemented in the theory by
defining Dirac brackets and this may be performed iteratively. We find
$(n,m\neq 0)$
\begin{equation}
\{\Phi_n,p_0\} = 0 \hspace{3cm} \{\Phi_n,\Phi_m\} = -2ik_n\delta_{m+n,0}\; ,
\end{equation}

\begin{equation}
\{\Phi_n,\beta \} = \{p_n,\beta \} = -\frac{1}{\sqrt{L}}\int_{-L/2}^{L/2}
dx\left[V''(\phi )-V''\left(\frac{q_0}{\sqrt{L}}\right)\right]e^{-ik_nx}\equiv
-\frac{\alpha_n}{\sqrt{L}}\; ,
\end{equation}
\begin{equation}
\{p_0,\beta \} = -\frac{1}{\sqrt{L}}\int_{-L/2}^{L/2}dx\; V''(\phi ) \equiv
-\frac{\alpha}{\sqrt{L}}\; ,
\end{equation}
\begin{equation}
\{p_0,p_0\} = \{\beta ,\beta \} = 0
\end{equation}

We implement first the pair of constraints $p_0\approx 0,\; \beta \approx
0$. The Dirac bracket $\{ \}^{\star}$ with respect to them is easily
constructed
\begin{equation}
\{f,g\}^{\star} = \{f,g\}-[\{f,p_0\}\{\beta ,g\}-(p_0\leftrightarrow \beta )]
\frac{\alpha}{\sqrt{L}}^{-1}.
\end{equation}
We may then set $p_0=0$ and $\beta =0$ as strong relations since, for example,
$\{f,p_0\}^{\ast}
= \{f,\beta \}=0$ for any arbitrary functional $f$ of our canonical variables.
The variable $p_0$ is thus removed from the theory. We conclude easily by
inspection that the brackets $\{ \}^{\ast}$ of the surviving canonical
variables coincide with the standard Poisson brackets except for the ones
involving $q_0$ and $p_n (n\neq 0)$
\begin{equation}
\{q_0,p_n\}^{\ast} = \{q_0,\Phi_n\}^{\ast} = -(\alpha^{-1}\alpha_n)
\end{equation}
For the potential given just after eq. (1) above we find
\begin{equation}
\{q_0,p_n\}^{\ast} = \{q_0,\Phi_n\}^{\ast} = -\frac{3\lambda
[2q_0q_{-n}+\int_{-L/2}^{L/2}
dx\varphi^2\; e^{-ik_nx}]}{[3\lambda (q_0/\sqrt{L})^2-m^2]L+3\lambda
\int_{-L/2}^{L/2} dx\varphi^2}
\end{equation}

Next we implement the remaining constraints $\Phi_n\approx 0\;(n\neq 0)$. We
find
\begin{equation}
C_{nm} = \{\Phi_n,\Phi_m\}^{\ast} = -2ik_n\delta_{n+m,0}
\end{equation}
and its inverse is given by $C^{-1}_{~~~nm}=(1/2ik_n)\delta_{n+m,0}$. The {\em
final} Dirac bracket which takes care of all the constraints of the theory is
then given by
\begin{equation}
\{f,g\}_D = \{f,g\}^{\ast} - \sum_n{}'\;
\frac{1}{2ik_n}\{f,\Phi_n\}^{\ast}\{\Phi_{-n},g\}^{\ast}.
\end{equation}
Inside this final bracket all the constraints may be treated as strong
relations and we may now in addition write $p_n=ik_nq_{-n}$. It is
straightforward
to show that
\begin{equation}
\{q_0,q_0\}_D=0 \hspace{0.5cm}
\{q_0,p_n\}_D=\{q_0,ik_nq_{-n}\}_D=\frac{1}{2}\{q_0,p_n\}^{\ast},
\hspace{0.5cm} \{q_n,p_m\}_D=\frac{1}{2}\delta_{nm}.
\end{equation}
It is also convenient to introduce the field $\pi (\tau ,x)$
\begin{equation}
\pi(\tau ,x) \equiv \varphi '(x) = \sum_n{}' \frac{p_n}{\sqrt{L}}e^{ik_nx}
\end{equation}
which like $\varphi$ is summed over the nonzero modes.

In order to remove the spurious finite volume effects in discretized
formulation we must take the {\em continuum limit} $L\rightarrow \infty$. We
have
as usual: $\Delta =2(\pi /L)\rightarrow dk,\; k_n=n\Delta \rightarrow k, \;
\sqrt{L}
q_{-n}\rightarrow \lim_{L\rightarrow \infty} \int_{-L/2}^{L/2}dx\varphi
(x)e^{ik_nx} \equiv \int_{-\infty}^{\infty}dx\varphi (x)e^{ikx}=\sqrt{2\pi}
\stackrel{\sim}{\varphi}(k)$ for $n,k\neq 0$. The sum over the nonzero mode in
(3) goes over to $\sqrt{2\pi}\varphi
(x)=\int_{-\infty}^{\infty}dk\stackrel{\sim}{\varphi}
(k)e^{-ikx}$ along with the restriction $\int_{-\infty}^{\infty}dx\varphi (x)=
\sqrt{2\pi}\varphi (0)=0$. Since the zero mode gives rise to the vacuum
expectation value of the quantized field $\phi$ it is clear that
$(q_0/\sqrt{L})
\rightarrow \omega (\tau )$. From
$\{\sqrt{L}q_m,\sqrt{L}q_{-n}\}_D=L\delta_{nm}/(2ik_n)$
following from the Dirac bracket between $q_m$ and $p_n$ for $n,m\neq 0$ in
(19) we derive, on using $L\delta_{nm}\rightarrow \lim_{L\rightarrow
\infty}\int_{-L/2}^{L/2}
dxe^{i(k_n-k_m)x}=\int_{-\infty}^{\infty}dxe^{i(k-k')x}=2\pi \delta (k-k')$,
that
\begin{equation}
\{\stackrel{\sim}{\varphi}(k),\stackrel{\sim}{\varphi}(-k')\}_D = \frac{1}{2ik}
\delta (k-k')
\end{equation}
where $k,k'\neq 0$. On making use of the integral representation of the sgn
function, $\in (x)=(i/\pi ){\cal P}\int_{-\infty}^{\infty}(dk/k)e^{-ikx}$ we
are led to the light-front Dirac brackets for the field $\varphi$
\begin{equation}
\{\varphi (x),\varphi (y)\}_D = -\frac{1}{4}\in (x-y)
\end{equation}
{}From (15) and (18) we derive similarly $\{\omega ,\omega \}_D=0$ and
\begin{equation}
\{\omega ,\varphi (x)\}_D = -\left(\frac{3\lambda}{4}\right) lim_{L\rightarrow
\infty}
\frac{\int_{-\infty}^{\infty}dy \in (x-y)[2\omega \varphi (y)+\varphi (y)^2]}
{(3\lambda \omega^2-m^2)L+3\lambda \int_{-\infty}^{\infty}dx\varphi (x)^2}
\end{equation}
We find that at the classical level $\{\omega ,\varphi (x)\}_D=0$ only in the
continuum limit and if the values of $\omega$ are such that the coefficient of
$L$ in the denominator of (22), e.g. $V''(\omega )$, is nonvanishing.

The resulting Dirac bracket (21) of $\varphi$ is the usual light-front
\cite{J}. However, in the interacting theory we must take into account also of
the implications of the {\em constraint} $\beta =0$. Its explicit form for the
potential under consideration is
\begin{equation}
L\left(\frac{q_0}{\sqrt{L}}\right)\left[\lambda
\left(\frac{q_0}{\sqrt{L}}\right)^2-m^2\right]
+ \lambda \int_{-L/2}^{L/2}
dx\left[3\left(\frac{q_0}{\sqrt{L}}\right)\varphi^2+\varphi^3\right]=0
\end{equation}
which goes over to
\begin{equation}
\omega (\lambda \omega^2-m^2) + \lambda \; lim_{L\rightarrow \infty}\frac{1}{L}
\int_{-L/2}^{L/2} dx[3\omega \varphi^2+\varphi^3]=0.
\end{equation}
At the tree level if $\omega$ is finite we find $V'(\omega )=\omega (\lambda
\omega^2-m^2)=0$
which gives rise to $\omega =0$ for the symmetric phase while $\omega =\pm
(m/\sqrt{\lambda})$ for the asymmetric phases. For the free field theory or
when we have the correct sign for the mass term (e.g., $m^2\rightarrow -m^2)$
in the interacting case we find $\omega =0$. The coefficient of $L$ in the
denominator of (22) is non-vanishing for these values for $\omega$ and
consequently the right hand side of (22) vanishes for these cases on removing
the finite size effects by taking the continuum limit. The {\em final}
Hamiltonian coincides with $H_c$
\begin{equation}
H = H_c = \int_{-\infty}^{\infty} dx\left[\frac{1}{2}(3\lambda
\omega^2-m^2)\varphi^2 + \lambda (\omega \varphi^3+\frac{1}{4}\varphi^4)
+ \frac{1}{4\lambda}(\lambda \omega^2-m^2)^2\right],
\end{equation}
and the Lagrange equations of motion are recovered from the Hamiltons equations
assuring
the self-consistency \cite{Di} of the procedure. It is clear from (24) and (25)
that the elimination of $\omega$ using the constraint would lead to a very
involved Hamiltonian except in the case we ignore the background field
altogether. It is clear also that all these results follow {\em immediately} if
we had worked directly in the continuum and interpreted $\delta (0)=L/(2\pi )$
with $L\rightarrow \infty$.

\vspace{1cm}
{\bf 3.} The quantized theory is obtained by the correspondence
$i\{f,g\}_D\rightarrow [f,g]$ where the quantities inside the commutator are
the corresponding quantized operators. In the interacting theory the operator
$\omega$ is seen to commute with itself and with the nonzero modes only in the
continuum limit and we are left with the {\em nonlinear operator constraint}
together with the Hamiltonian operator. The higher order corrections to (24) in
the renormalized field theory will alter the tree level values of $\omega$
since, we do {\em not} have any physical considerations to normal order the
constraint equation. The commutation relations of $\varphi$ may be realized in
momentum space through the expansion $(\tau =0)$
\begin{equation}
\varphi (x) = \frac{1}{\sqrt{2\pi}} \int_0^{\infty} \frac{\theta
(k)}{\sqrt{2k}}
[a(k)e^{-ikx}+a^{\dag}(k)e^{ikx}]
\end{equation}
\begin{sloppypar}
\noindent
where $a(k)$ and $a^{\dag}(k)$ satisfy the canonical commutation relations,
viz, $[a(k),a(k')^{\dag}]=\delta (k-k'), [a(k),a(k')]=0$, and
$[a(k)^{\dag},a(k')^{\dag}]=0$ while the $\omega$ commutes with them and thus
is proportional to the identity operator in the Fock-space. The vacuum state is
defined by $a(k)|vac\rangle =0, k>0$. The longitudinal momentum operator is
$P^+
=\int dx:\varphi^{'2}:$ and the light-front energy is $P^-=H=\int dx:V(\phi ):$
where we normal order with respect to the creation and destruction operators to
drop unphysical infinities and we find $[a(k),P^+]=ka(k),
[a^{\dag}(k),P^+]=-ka^{\dag}(k)$.
The values of $\omega =\langle |\phi |\rangle_{vac}$ obtained from solving
$V'(\omega )=0$ {\em characterize} the (non-perturbative) vacua and the Fock
space is built by applying the nonzero mode operators on the corresponding
vacuum state. The reflection symmetry $\phi \rightarrow -\phi$ is broken
spontaneously when $\omega \neq 0$. We remark that in view of the constraint
(24) and the non-commutability of $\omega$ with the operators $a(k)$ and
$a^{\dag}(k)$ in finite volume a nonperturbative computation in the DLCQ would
be much more difficult to handle than in the continuum formulation except for
in the usually considered case where the background field is ignored
altogether. The high order quantum corrections would alter the value of
$\omega$ as determined from the renormalized constraint equation and it may be
shown that we do obtain significant deviations \cite{In} from the well known
``duality'' relation \cite{Col}. It is worth remarking that in the present case
the zero mode (of phi) is the background field and is essentially a c-number
over the Fock space. This is in contrast to the case of the light-front
quantization of the bosonic version of the Schwinger model obtained by
functionally integrating over the fermions and introducing the scalar field to
obtain a local field theory. Here in order to ensure at the quantum level the
symmetry of the Lagrangian with respect to the shift by a constant of the
scalar field (chiral symmetry), a zero mode from the only other field
available, viz, the gauge field, must be an operator and canonically conjugate
to the zero mode operator of the scalar field. This model discussed by modified
procedure \cite{T} can also be handled by following the standard Dirac
procedure.
\end{sloppypar}

\vspace{1cm}
{\bf 4.} We conclude thus that in view of the constraint in the light-front
quantized field theory we do obtain a description of the spontaneous symmetry
breaking. The treatment of the non-perturbative effects in the DLCQ, contrary
to the common understanding, seems to be quite a difficult task in view of the
non-commutability of the background field with the nonzero modes and the
presence of the non-linear constraint. The continuum formulation is more
convinient in that it can be renormalized straightforwardly and high order
corrections computed from the renormalized constraint equation.

\vspace{1cm}
\noindent
{\bf Acknowledgements}

The author acknowledges the hospitality of the Department of Physics at the
Ohio State University and University of Padova and financial grants from the
INFN, Padova and the CNPq of Brasil. The author is grateful to Steve Pinsky for
informing him on the motivations for this work. He acknowledges with thanks
constructive suggestions from him and Avaroth Harindranath, who helped him much
in getting acquainted with the subject and for his updated {\em ``Sources for
Light Front Physics''}. I am indebted to Robert Perry for his continued
interest and stimulating suggestions helping toward the completion of the work.
Comments from Junko Shigemitsu, Kent Hornbostel, Stan Glazek, Katsumi Tanaka,
and Stuart Raby are also acknowledged.


\newpage
\begin{thebibliography}{100}
\bibitem{Dirac}P.A.M. Dirac, Rev. Mod. Phys. {\bf 21} (1949) 392.
\bibitem{Wein} S. Weinberg, Phys. Rev. {\bf 150} (1966) 1313.
\bibitem{Pauli} H.C. Pauli and S.J. Brodsky, Phys. Rev. {\bf D~ 32} (1985) 1993
and
2001; S.J. Brodsky and G.P. Lepage, in {\em Perturbative Quantum
Chromodynamics}, edited by A.H. Mueller, World Scientific, Singapore, 1989. In
our context see the {\em recent reviews}: S.J. Brodsky and H.C. Pauli, SLAC
preprint SLAC-PUB-5558/91 and K. Hornbostel, Cornell University preprint, CLNS
91/1078.
\bibitem{Wil} K.G. Wilson, in {\em Lattice '89}, Proceedings of the
International Symposium, Capri, Italy, 1989, edited by R. Petronzio et al.
[Nucl. Phys. B (proc. Suppl.)] {\bf 17} (1990); R.J. Perry, A. Harindranath,
and K.G. Wilson, Phys. Rev. Lett. {\bf 65} (1990) 2959.
\bibitem{Mus} D. Mustaki, S. Pinsky, J. Shigemitsu and K. Wilson, Phys. Rev. D
{\bf 43} (1991) 3411.
\bibitem{Chang} S.J. Chang, Phys. Rev. {\bf D~ 13} (1974) 2778; A. Harindranath
and
J.P. Vary, Phys. Rev. {\bf D~ 36} (1987) 1141.
\bibitem{Col} S. Coleman and E. Weinberg, Phys. Rev. {\bf D~ 7} (1973) 1888; B.
Simon and R.B. Griffiths, Comm. Math. Phys. {\bf 33} (1973) 145.
\bibitem{Di} P.A.M. Dirac, Canad. J. Math. {\bf 1} (1950) 1; {\em Lectures in
Quantum Mechanics}, Benjamin, New York, 1964; E.C.G. Sudarshan and N. Mukunda,
{\em Classical Dynamics: a modern perspective}, Wiley, N.Y., 1974.
\bibitem{Mask} T. Maskawa and K. Yamawaki, Prog. Theor. Phys. {\bf 56} (1976)
270; R.S. Wittman, in Nuclear and Particle Physics on the Light-cone, eds. M.B.
Johnson and L.S. Kisslinger, World Scientific, Singapore, 1989.
\bibitem{Th} Th. Heinzl, St. Krusche, and E. Werner, Regensburg preprint TPR
91-23, Phys. Lett. {\bf B~ 272} (1991) 54.
\bibitem{S} S. Schlieder and E. Seiler, Commun. Math. Phys. {\bf 25} (1972) 62.
\bibitem{J} S.J. Chang, R.G. Root and T.M. Yang, Phys. Rev. {\bf D~ 7} (1973)
1133.
\bibitem{In} {\em in preparation}.
\bibitem{T} Th. Heinzl, St. Krusche and E. Werner, Phys. Lett. {\bf B~ 256}
(1991) 55.
\end{thebibliography}
\end{document}